\documentclass[a4paper,12pt]{article} % Set paper size and document type
\usepackage{lmodern} % Use a slightly nicer looking font
\usepackage{url} % Proper formatting for URLs
\usepackage{graphicx} % Handle inclusion of non-PDF graphics
\usepackage{subfig} % Allow sub-figures inside a figure
\usepackage{enumitem} % Allow lists to pick up numbering where the last list left off
\usepackage{nccmath}
\usepackage{csquotes}
\usepackage{amsmath}
\usepackage{amsthm}
\usepackage{amsmath}
\usepackage{amssymb}
\usepackage{mathtools}
\usepackage{lipsum}  
\usepackage{algorithm}
\usepackage{algorithmic}
\usepackage[
    backend=biber,
    natbib=true,
    style=numeric,
    doi=true,
]{biblatex}
\usepackage[hidelinks, pdfusetitle]{hyperref}

\newtheorem*{definition}{Definition}

\newtheorem{proposition}{Proposition}
\newtheorem{theorem}{Theorem}
\usepackage[toc,page]{appendix}

\usepackage[margin=30mm]{geometry}

\usepackage{listings}
\usepackage{color}

\definecolor{mygreen}{rgb}{0,0.6,0}
\definecolor{mygray}{rgb}{0.5,0.5,0.5}
\definecolor{mymauve}{rgb}{0.58,0,0.82}

\title{\textbf{RIVCoin: an alternative, integrated, CeFi/DeFi-Vaulted Cryptocurrency}}
\author{
  Roberto Rivera\\
  \texttt{rob@riv-capital.com}
  \and
  Guido Rocco\\
  \texttt{grk@riv-capital.com}
  \and
  Massimiliano Marzo\\
  \texttt{max@riv-capital.com}
  \and
  Enrico Talin\\
  \texttt{eta@riv-capital.com}
  \and
  Ammar Elsabe\\
  \texttt{ammar.elsabe@riv-capital.com}
}

\begin{document}

\maketitle

\begin{abstract}
This whitepaper introduces RIV Coin, a cryptocurrency that is fully stabilized by a diversified portfolio of invested reserves that are a) evaluated by professional independent third parties, and b) auditable and provable by the protocol.
It is born and managed as a decentralized token, minted by a Decentralized Autonomous Organization (DAO).
All wealthier Users are then accepting a redistribution of income, to the benefit of those who have purchased less tokens.
In cooperative Game Theory, maximization of the economic benefit of the ecosystem is achieved when players' incentives are perfectly aligned.
The proposed model allows for alignment of incentives: decreasing the risk exposure by wealthier Users, but implicitly increasing that of smaller ones to a level perceived by them as still sustainable and never creating ultra-speculative positions. In other words, wealthier Users stabilize the risk associated with the market value of portfolios in which the reserves are invested in Centralized and Decentralized Finance, without falling into the “bet scheme”.
Users indirectly benefit from the access to the rewards of sophisticated cryptocurrency portfolios hitherto precluded to them, as well as having access to a real redistribution of wealth, without this turning into a disadvantage for the wealthy User, who benefits from the greater stability created by the huge influx of smaller Users.
Therefore, the progressive growth becomes additional value that tends to stabilize over time, optimizing RIV Coin on the systemic risk level. 
\end{abstract}
\pagebreak

\section{Introduction}

In \citeyear{Satoshi2008}, \citeauthor{Satoshi2008} introduced Bitcoin \cite{Satoshi2008} as a decentralized solution for reducing fraud in electronic payments by removing reliance on trusted third parties through cryptographic techniques. While Bitcoin succeeded in creating a trustless transaction system, it fell short of establishing a stable medium of exchange. Its rigid, algorithmic monetary policy, though resilient against manipulation, left it vulnerable to extreme volatility, limiting its effectiveness as a true currency.

The broader decentralized finance (DeFi) ecosystem has since struggled with this core challenge: reconciling decentralization with economic stability. While cryptocurrencies have innovated in removing custodial trust, they have not yet realized a system capable of adapting dynamically to economic realities without reintroducing centralized points of failure.

Meanwhile, traditional fiat currencies achieve relative stability through active monetary interventions--central banks managing interest rates and liquidity to stabilize economies. However, such systems are fundamentally dependent on perpetual debt expansion, introducing long-term risks of monetary devaluation and systemic fragility.

Despite their structural differences, both fiat and decentralized currencies ultimately derive value from trust: one rooted in institutions, the other in algorithms. Yet neither has fully solved the problem of creating a truly stable, decentralized medium of exchange anchored in real, productive value.

In the following sections, we explore the emergence of stablecoins as a bridge solution to these challenges, and examine why a more fundamental shift is needed.

\subsection{Background}

In response to the volatility of early cryptocurrencies, fiat-pegged stablecoins emerged as an attempt to introduce price stability into the digital asset ecosystem.
By anchoring digital tokens to sovereign currencies like the US Dollar, stablecoins such as Tether and USDC sought to create a reliable medium of exchange within crypto markets.
According to the Bank for International Settlements (BIS) in October 2019 in its ``Investigating the impact of global stablecoins'' \cite{bis}

\begin{displayquote}
``The first wave of cryptoassets, of which Bitcoin is the best known, have so far failed to provide a reliable and attractive means of payment or store of value. This reflected the preference of main market participants to base transactions and payments on sovereign fiat currencies, particularly the US Dollar.''
\end{displayquote}

Various taxonomies of stablecoins have been proposed in the literature. For instance, G. Liao and J. Carmichael \cite{liao} classify stablecoins into:
\begin{itemize}
    \item \textbf{Public Reserve-Backed Stablecoins}, such as Tether, USDC, and Paxos Dollar;
    \item \textbf{Public Algorithmic Stablecoins}, such as DAI and FRAX;
    \item \textbf{Private Stablecoins}, such as JPM Coin.
\end{itemize}

This classification highlights two fundamental axes: the \textbf{public versus private} nature of issuance, and the \textbf{mechanism of value stabilization}--whether through hard collateralization or algorithmic control. Each model involves trade-offs among decentralization, resilience, and regulatory exposure. 

Public reserve-backed stablecoins depend on centralized custodians for collateral management, introducing custodial and regulatory risk. Algorithmic stablecoins, in contrast, attempt to maintain pegs through supply manipulation, but remain highly vulnerable to market shocks, as seen in the collapse of TerraUSD.

\subsection{Toward a Reserve-Backed Digital Economy}
Although stablecoins have provided a degree of price stability within crypto markets, they inherit many of the structural vulnerabilities of the fiat systems they mirror. Both fiat currencies and stablecoins ultimately lack a key property that underpinned historical monetary systems like the gold standard: \textbf{intrinsic reserve backing}. Under the gold standard, monetary value was directly tied to verifiable, tangible assets. The physical limitation of gold supply enforced fiscal discipline, anchoring currency stability across generations.

Modern fiat currencies are primarily supported by debt issuance and dynamic monetary policy, rather than by tangible reserves.
Central banks manage fiat stability through interest rates and open market operations, aiming to control inflation and promote economic growth.
They are, in essence, \textit{algorithmic stablecoins underwritten by economic output and sovereign enforcement}, rendering them more resilient, yet fundamentally prone to the same structural instabilities.
However, this system inherently depends on the perpetual expansion of liabilities -- governments continually issue new debt to finance spending, while the money supply grows to accommodate these obligations. Over time, this dynamic has led to cycles of inflation, monetary devaluation, and systemic financial fragility.

For a currency, digital or otherwise, to serve as a true store of value and unit of account, it must break free from debt-dependence and/or artificial pegs. It must be supported instead by a growing, diversified base of real reserves that reflect genuine productive value.

Stablecoins have made important contributions by bridging fiat and crypto markets, but they are transitional instruments rather than final solutions. We believe the next evolution requires returning to first principles: a monetary system where value is directly anchored in real-world assets, professionally managed and dynamically adapted to changing economic conditions, but without relying on perpetual debt expansion.

This is the philosophy behind \textbf{RIV Coin}: a currency backed by an actively managed treasury (``The Vault''), composed of tangible, productive assets, rather than debt instruments or unsupported issuance mechanisms. Through Proof of Integrity, regular audits, and a sustainable treasury strategy, RIV Coin aims to provide a reliable, scalable, and inflation-resilient monetary alternative for the digital age.

\section{RIV Coin, a reserve backed cryptocurrency}
Our purpose is to create a cryptocurrency, with emphasis on \textit{currency}, a word which gets its origin ``from Medieval Latin currentia, literally: a flowing, from Latin currere to run, flow'', and literally meaning ``medium of exchange'' \cite{currencydefinition}.

We aim to develop a stable medium of exchange backed by an actively managed reserve of real assets, combining transparent, professional treasury management with decentralized security models.

We will accomplish this through the introduction of RIV Coin, a native Layer 1 cryptocurrency backed by a diversified reserve portfolio--hereinafter referred to as the Vault--which holds both traditional financial (CeFi) assets and decentralized finance (DeFi) positions.
In the system we are building, the Vault guarantees the total number of coins minted and, as a last resort, it can always be sold to reimburse part/all the coins in circulation.
the Vault can fluctuate in its value, but it cannot be used for anything else except for guaranteeing the minimum intrinsic value of all the coins minted.

The Vault's CeFi holdings are initially periodically, then in real time, audited and notarized by professional independent third parties (typically fund administrators and/or auditors), while the DeFi positions remain transparently verifiable on-chain. The combined portfolio undergoes continuous proof generation, ensuring that the system can attest to its Net Asset Value (NAV) without exposing sensitive operational data. This structure fosters verifiable trust, protects against information asymmetry, and mitigates exploitative practices such as front-running.

RIV Chain, the native network supporting RIV Coin, utilizes a Proof-of-Stake consensus mechanism in which validators and their delegators secure the chain and earn yield proportional to the Vault's growth.

The seignorage, defined as the return generated by the reserves pledged within the Vault, will be used for a democratic distribution of wealth, subject to explicit actions to be implemented by the Users. 
In contrast to traditional economies where seigniorage often drives ``bad inflation'' (as it makes governments to be self confident and print money beyond real growth) through uncontrolled monetary expansion, RIV Coin's model redistributes the return transparently and sustainably.
Growth from the underlying assets flows back to validators and their delegators, aligning incentives while promoting ``good inflation'' -- an increase in value directly linked to productive asset management rather than unchecked monetary supply.

\subsection{Growing Reserves}

The long-term stability of RIV Coin is underpinned by the Vault: a diversified reserve invested in actively managed traditional (CeFi) and decentralized finance (DeFi) assets.
The Vault is designed not merely to protect the value of RIV Coin, but to grow it sustainably over time through long-term investment strategies, low turnover, high risk-adjusted returns.

Reserves are continuously expanded through two primary mechanisms:
\begin{itemize}
    \item \textbf{Primary Reserve Growth}: New capital inflows are generated through non-inflationary minting mechanisms, where newly issued RIV Coins are backed 1:1 by fresh fiat or crypto assets pledged into the Vault.
    \item \textbf{Organic Investment Returns}: the Vault's assets are strategically deployed across diversified financial products--in a long-term, granular asset allocation. This produces organic return that grows the reserve base without requiring new issuance.
\end{itemize}

To safeguard the integrity and sustainable expansion of reserves, the Vault employs several critical design principles:
\begin{itemize}
    \item \textbf{Zero-Knowledge Transparency}: the Vault's Net Asset Value (NAV) is periodically validated through zero-knowledge proofs, maintaining performance transparency without revealing sensitive portfolio details. This structure protects against front-running and other exploitative behaviors.
    \item \textbf{Independent Auditing}: Regular third-party audits independently verify the composition and value of the Vault, ensuring user confidence in the reserve's true backing.
    \item \textbf{Resilience through Diversification}: Assets are granularly diversified across sectors and geographies to minimize systemic risk. Even in the event of a black swan market disruption, the impact on the Vault's overall value is highly mitigated, and the overshooting effects are neutralized in the short run.
\end{itemize}

Importantly, all reserves are pledged exclusively in favor of tokenholders, ensuring that RIV Coin remains a non-custodial asset without exposure to rehypothecation or leverage risks beyond the holder's control. Intrinsic value per RIV Coin is calculated as the total value of the Vault divided by the total circulating supply, creating a verifiable ``value in last resort'' for every user.

Through this dual engine of secured inflows and organic investment growth, RIV Coin's reserves are structured not merely to maintain stability, but to steadily expand over time--fortifying the ecosystem’s long-term value proposition.

\section{Theoretical Foundation} 
Our solution rests on the foundational insights from cooperative game theory, which illustrate that expanding a market to a broader pool of participants enhances both price discovery and overall market stability.
In simpler terms, when a greater number of investors take part in trading a particular security, the collective information they bring becomes reflected in market prices more efficiently.
Private information quickly transitions into the public domain, improving transparency and liquidity while dampening volatility. 

From a theoretical perspective, this improved informational efficiency leads to more accurate and stable valuations.
The trade-off, however, is that returns may be slightly lower than in highly concentrated markets that reward a smaller pool of well-informed participants.
Yet, under cooperative game theory, agents are collectively better off accepting a modest reduction in returns in exchange for higher-quality information, reduced volatility, and deeper liquidity.
The expanded participation effectively raises the “floor” for market stability and precision in pricing. 

While wealth redistribution has historically relied on fiscal tools like taxation, RIVCoin introduces a disruptive alternative: decentralized redistribution through network security.
By rewarding Users who support network security, RIVCoin enables wealth to flow through participation, not policy. 

In the remainder of this section, we will delve into the theoretical underpinnings of our proposal, exploring the role of information, noise, and equilibrium in financial markets.

\subsection{Information, Noise and Equilibrium}

In what follows we use some crucial results known in the literature to prove that an
higher number of informed and uninformed traders contributes to informational efficiency
of the market, by making prices of any security less volatile.
To do so, we take advantage of some results contained in several contributions in a well consolidated literature. We reinterpret the results obtained by \cite{kyle89} \cite{rindi08} \cite{djrindi}, to focus on the role of information diffusion, when agents can have differential information about the price of single risky security.

As a preview of our results, we show that: 
\begin{enumerate}
	\item[1.] In a model with rational expectations, but with agents having differential information, increasing the number of agents (uninformed and informed) \textbf{improves informational efficiency and lowers price volatility. }
	\item[2.] \textbf{Increasing the number of both informed and uninformed agents will reduce the negative impact played by noise traders:} the return will be lower, as a result of the improved ability of prices to convey information to the market, but the volatility will be lower as well. 
	\item[3.] In Fully Revealing Rational Expectations equilibrium, when the number of agents (both informed and uninformed) tends to be very large, private information becomes public, since uninformed agents learn about the private signal owned by informed agents. 
\end{enumerate} 
As a general comment, increasing the number of agents improves the quality of the market as a whole, lowering price volatility and increasing the price level, converging to the equilibrium level. In a situation characterized by higher informational efficiency, security prices tend to increase. In this context, the role of both the fundamentals of the asset and its supply becomes crucial in order to balance the attractiveness of the security and to guarantee the access to a large number of individual investors. The important point to be stressed is that more agent buy the asset, higher is its price and lower its volatility for a fixed asset supply. 

\subsection{The Model}

Let us start with designing a market with asymmetric information. Assume there exists
two groups of agents, both are assumed to be risk-averse: $N$ are assumed to be informed
and $M$  are assumed to be uninformed. Intuitively, we can imagine that informed agents
are professional traders and uninformed agents are non-professional traders,
considering trading activity almost like a spare-time job. Furthermore, we assume the
presence of noise traders or agents in measure equal to $Z$: these individuals are assumed to be
interpreted like automated traders.

The utility function for both informed and non-informed traders can be represented
by following mean-variance utility: 
\begin{equation}\label{uti}
	U = E (\tilde{W}) - \frac{\alpha}{2} Var(\tilde{W})
\end{equation}
where $ \alpha  $  is the risk aversion coefficient and $ \tilde{W} $ is the end of period wealth of each agent given by:
\begin{equation}\label{w}
	\tilde{W} = Y ( \tilde{S} - p )
\end{equation}
where $Y$ is the demand of each agent, $ p $ is the today's price of the security, and $ \tilde{S} $ is the future’s price of the security, which is assumed to be a random variable, with an i.i.d.
normally distributed probability density function:  
\begin{equation}\label{S}
  f(\tilde{S} ) \sim N(S, {\sigma_S}^2  )
\end{equation}
where, of course, $ E(\tilde{S} ) = S $, $ Var(\tilde{S} ) = \sigma^{2}_{S} $.

For simplicity, we assume that each agent has an initial endowment of risky asset equal to zero, and there is not a risk-free security. These assumptions are non-crucial to our results, but help us to save the burden of notation.

By maximizing utility (\ref{uti}) subjected to wealth accumulation constraint (\ref{w}), we obtain the following demand of risky asset expressed by:
\begin{equation}\label{Y}
	Y = \frac{E( \tilde{S}) - p  }{  \alpha Var( \tilde{S})}
\end{equation}
Equation (\ref{Y}) is the demand expressed by each agent who does not have any private information: she makes a forecast about the future evolution of the security price $ \tilde{S} $. Each agent expects to gain a return measured by the difference from future price $ \tilde{S} $ and current
price $ p $: $ \tilde{R} = \tilde{S} - p $.

The demand is weighted by the volatility of the risky asset, measured by
its variance, $ Var(\tilde{S}) = \sigma^{2}_{S} $ and by risk aversion coefficient $ \alpha $: higher risk aversion, and higher variance, lower will be demand for the risky security.

Furthermore, we assume that informed trader, here assumed to be of mass $ N $, receive
a signal $ \tilde{X} $ at time $ t=0 $ about  the behavior of future asset $\tilde{S}  $. 
The signal is private, proprietary information of only informed traders.

There can be many examples of such situation: we can imagine, for example, that the issuance of a digital coin can be reserved to only few, selected investors. Therefore, the question we need to answer is whether extending this possibility to many other non-informed traders can improve
the informational quality of the market and reduce the volatility of the asset.

The private signal received by informed traders is assumed to be a random variable
conditional to the evolution of future security's price $ \tilde{X} \vert \tilde{S} $, with a probability density function i.i.d. and normal: $ f (\tilde{X} \vert \tilde{S})   $: 
\begin{equation}\label{fcond}
	f (\tilde{X} \vert \tilde{S}) \sim N(\tilde{S}, \sigma^{2}_{X}  )
\end{equation}
Clearly, the expected value of the signal is exactly equal to the future value of the security:
on average the signal is a non-distorted forecast of the future price. The
informed agent makes her decisions based on information coming from the
signal given the probability density (\ref{fcond}).

Therefore, the demand of informed agent $ Y_{I} $ is expressed as follows:
\begin{equation}\label{Yi}
	Y_{I} = \frac{E(  \tilde{S} \vert \tilde{X}  ) - p  }{  \alpha Var( \tilde{S} \vert \tilde{X})}	
\end{equation}
As a final step, we explicit the role of noise traders, whose demand function is assumed
to be a pure random variable $ \tilde{h} $ is normally distributed and i.i.d., such that:
\begin{equation}\label{h}
	f(\tilde{h}) \sim  N(0, \sigma^{2}_{h})
\end{equation}
To find equilibrium level of prices, we need to obtain an analytical expression for the
statistics $ E(  \tilde{S} \vert \tilde{X}  ) $ and $ Var( \tilde{S} \vert \tilde{X}) $ included in \ref{Yi} . To this purpose, we can take advantage of the results contained in the following proposition.
\begin{proposition} \label{pro1}
	Given assumptions (\ref{S}) and (\ref{fcond}), the distribution of the future asset's price $ \tilde{S} $, conditional to the signal $ \tilde{X} $, is i.i.d. and normal, given by: 
	\begin{equation}\label{ScondX}
		f(\tilde{S} \vert \tilde{X}
		) \sim N (\mu_{n}, \sigma_{n}^{2}) 
	\end{equation}
where: 
\begin{equation}\label{mun}
	\mu_{n} = \frac{ \lambda_{S} S + \lambda_{X} \tilde{X} }{\lambda_{S} + \lambda_{X}}; \qquad \sigma_{n}^{2} = \frac{ 1}{\lambda_{S} + \lambda_{X}}
\end{equation}
and: 
\begin{equation}\label{preci}
	\lambda_{S} = \frac{1}{\sigma_{S}^{2}}; \qquad \lambda_{X} = \frac{1}{\sigma_{X}^{2}} 
\end{equation}
 
\end{proposition}

where $ \lambda_{S} $, $ \lambda_{X}  $ are the precision, defined as the inverse of the variances.
\begin{proof}
  See Appendix \ref{appB}.
\end{proof}

Now, we can easily rewrite the distribution in (\ref{ScondX}) as follows: 
\begin{equation}\label{condi2}
	f(\tilde{S} \vert \tilde{X}	) \sim N ( \mu_{1}\tilde{S} + (1 - \mu_{1}) \tilde{X}  ,  (1 - \mu_{1})\sigma_{S}^{2}) 
\end{equation}
where $ \mu_{1} $ is defined as: 
\begin{equation}\label{mu1}
	\mu_{1} = \frac{ \lambda_{S}  }{\lambda_{S} + \lambda_{X}}
\end{equation}
Essentially, the result proved in Proposition (\ref{pro1}) delivers a Bayesian posterior for the distribution of the future assets' price conditional to the signal $ \tilde{X} $, as represented by equation (\ref{ScondX}), or (\ref{condi2}).  The prior distribution is given by (\ref{S}), the likelihood is identified with distribution in (\ref{fcond}).\

Now we have all the elements to define the equilibrium price. We distinguish two cases: a na\"{i}ve expectation equilibrium and a Rational Expectation Equilibrium. 
\subsubsection{Na\"{i}ve Expectations}
In this context, each trader submits her demand to the market where the price adjust in order to equate demand with supply. The market clearing condition is defined according to the following equilibrium condition: 
\begin{equation}\label{equili}
	N Y_{I} + M Y + Z \tilde{h} = 0
\end{equation}
Therefore, to obtain the equilibrium price level, we need to substitute the expression of the demand coming from informed trader $ Y_{I} $, non-informed traders $ Y $ and noise traders $ \tilde{h} $ into equilibrium (\ref{equili}).  The result is condensed in the following theorem. 
\begin{theorem} \label{th1}
	The equilibrium level of price of the risky asset is given by:
	\begin{equation}\label{pricena}
		p = \theta_{1} E (\tilde{S} \vert \tilde{X}) + (1 - \theta_{1}  ) E (\tilde{S}) + \theta_{2} \tilde{h}
	\end{equation}
where $  \theta_{1} $,  $ \theta_{2}$, are defined, respectively as: 
\begin{eqnarray}
	\theta_{1}  & = & \frac{ N Var( \tilde{S}  )}{ N Var( \tilde{S} \vert \tilde{X}  ) + M Var(\tilde{S}) } \label{theta1}  \\
	\theta_{2}  & = &  \frac{ \alpha Var( \tilde{S} \vert \tilde{X}  ) Var( \tilde{S}  )}{ N Var( \tilde{S} \vert \tilde{X}  ) + M Var(\tilde{S}) } \label{theta2}
\end{eqnarray}

\end{theorem}
\begin{proof}
  See Appendix \ref{appC}.
\end{proof}
According to the na\"{i}ve equilibrium outlined in Theorem \ref{th1}, each trader does not act strategically, in the sense that when submitting an order, she does not take into account the impact of her order on the equilibrium price or market equilibrium.

Despite the simplicity of this equilibrium, it is worth to check that, when the number of non-informed traders, represented, for example, by a set of large passive traders becomes arbitrarily large, i.e. when $ M \rightarrow \infty $, from (\ref{theta1}) and (\ref{theta2}) we obtain immediately that $ \theta_{1} \rightarrow 0 $, $ \theta_{2} \rightarrow 0 $, so that: $ p \rightarrow E(\tilde{S}) $: the increase of the number of passive traders makes the price to converge to their expectations of future price $ E(\tilde{S} )$.

If, on the other hand, is the number of informed traders to become arbitrarily large, when $ N \rightarrow \infty $, we obtain that $ \theta_{1} \rightarrow 1 $, $ \theta_{2} \rightarrow 0 $: the equilibrium price converges to the expectation formulated by informed agents.  

\textbf{Interestingly, even from this simple representation of the equilibrium, when the size of both type of agents tend to be very large, the role of noise trader tends to be neutralized: this is a simple example showing the increase in information efficiency.} The final equilibrium price converges to the expectations made by the relatively large group of traders, whose expectation becomes dominant in the market. 

It is important to stress the importance of this preliminary result: increasing the number of individuals participating in the market, neutralizes the noise trader demand, making the price level more informative and anchored to expectations formulated by market players, no matter if they are informed or uninformed. 

The equilibrium price is driven by the relative size of each group of individuals (informed, uninformed, noise traders). If non-informed traders become arbitrarily large, private information and noise trading activity both become irrelevant in price determination.  

We now extend this result to the context of a rational expectations equilibrium. 

\subsection{Rational Expectations Equilibrium}
In this context, we extend previous results to the context characterized by Rational Expectations Equilibrium.  Let us start by assuming that the signal is truly informative about the evolution of the future price of the asset. This implies that the relationship existing between the price $ \tilde{S} $ and the signal $ \tilde{X} $ can be represented according to the following equation: 
\begin{equation}\label{raS}
	\tilde{S} = \tilde{X} + \epsilon
\end{equation}
we assume now that the signal is normally distributed and i.i.d. with zero mean and constant variance $ \sigma_{X}^{2} $: $ \tilde{X} \sim N(X,\sigma_{X}^{2} ) $.  Moreover, we assume that the error  $ \epsilon $ is also i.i.d. normally distributed, with zero mean and constant variance $ \sigma_{\epsilon}^{2} $: $ \epsilon \sim N(0,\sigma_{\epsilon}^{2} ) $.  Moreover, the signal $\tilde{X}$ and the error $ \epsilon$ are assumed to be orthogonal. From these assumptions, we obtain that the stock price is normally distributed with the following density function: 
\begin{equation}\label{sdist}
	\tilde{S}  \sim N(0,\sigma_{X}^{2} + \sigma_{\epsilon}^{2} )
\end{equation}
Assume that informed traders (or insiders) observe the signal $ \tilde{X} $. Therefore, the conditional expectation and the variance of $ \tilde{S} $ are: 
\begin{equation}\label{cosa}
	E(\tilde{S} \vert \tilde{X}    ) = X; \qquad Var(\tilde{S} \vert \tilde{X}    ) = \sigma_{\epsilon}^{2}
\end{equation}
Each type of trader uses the actual price $ p $ as a mean to update their expectations about future price. Uninformed agents form their expectations by using only current price, so that: $ E(\tilde{S} \vert \tilde{X}    ) = X $ and $ Var(\tilde{S} \vert \tilde{X}    ) = \sigma_{\epsilon}^{2} $.  

On the other hand, informed agent adopt the signal together with the current price in order to  make expectations about future price, so that their expectation is $ E(\tilde{S} \vert \tilde{X},p) $.  If price is correlated with insiders' signal and with the true value of the security, we have that: $ E(\tilde{S} \vert \tilde{X},p) = E(\tilde{S} \vert \tilde{X}) $, and $ Var(\tilde{S} \vert \tilde{X},p) = Var(\tilde{S} \vert \tilde{X}) $.  

With these assumptions, the demand of risky security coming from non-informed, $ Y $, and from informed agents $ Y_{I} $ are given, respectively, by: 

\begin{eqnarray}
	Y & =  & \frac{ E(\tilde{S} \vert p ) - p   }{\alpha Var (\tilde{S} \vert p )   }  \label{dema}    \\  
  Y_{i} & =  & \frac{ E(\tilde{S} \vert  \tilde{X}   p ) - p   }{\alpha Var (\tilde{S} \vert \tilde{X} p )   }  =  \frac{ E(\tilde{S} \vert  \tilde{X} ) - p   }{\alpha Var (\tilde{S} \vert \tilde{X} )   }  \label{dema2}  
\end{eqnarray}
To get the equilibrium price, let us assume that uninformed agents conjecture the price to be a function of the signal and of the demand coming from noise traders according to the following function: 
\begin{equation}\label{p}
	p = \gamma_{1} \tilde{X} + \gamma_{2} \tilde{h}
\end{equation}
As discussed previously, informed agents make investment decisions by considering the role of signal which is assumed a sufficient statistic for the true price. Therefore, the demand of informed agent is: 
\begin{equation}\label{demai}
	 Y_{i}  =   \frac{\tilde{X} - p   }{\alpha \sigma_{\epsilon}^{2}   }
\end{equation}
Uninformed agents, instead, define their expectations for their demand conditional to actual price, as defined in equation (\ref{p}).  Therefore, by applying Projection Theorem (see Appendix \ref{appA}), we obtain the following expression for the conditional expectation $ E(\tilde{S} \vert p ) $: 
\begin{equation}\label{Sp}
	E(\tilde{S} \vert p ) = E(\tilde{S}) + \frac{Cov(\tilde{S},p)}{Var(p)} ( \tilde{S} - E(p))
\end{equation}
We need now to get an analytical expression for $ Cov(\tilde{S},p) $. To this purpose, using the expression given in (\ref{p}) for $ p $, together with the definition of $ \tilde{S} $ given in (\ref{raS}), we can write: 
\begin{eqnarray}
	Cov(\tilde{S},p) & =  &E ( \tilde{X} + \epsilon, p ) = E \left[ ( \tilde{X} + \epsilon)  ( \gamma_{1} \tilde{X} + \gamma_{2} \tilde{h} )    \right] =  \label{cova} \\
	& = & \gamma_{1} E( \tilde{X}  ) = \gamma_{1} \sigma_{X}^{2} \label{cova2}
\end{eqnarray}
The other ingredient which is needed is represented by the variance of the price: $ Var(p) = \gamma_{1}^{2} \sigma_{X}^{2} + \gamma_{2}^{2} \sigma_{h}^{2}  $. Therefore, the expectation of future stock's price conditional to today's price $ p $ will be: 

\begin{equation}\label{condS}
	E ( \tilde{S} \vert p) = \frac{ \gamma_{1} \sigma_{X}^{2} }{\gamma_{1}^{2} \sigma_{X}^{2} + \gamma_{2}^{2} \sigma_{h}^{2}    } p  = \theta p
\end{equation}

where $ \theta  $ is defined as: 

\begin{equation}\label{theta}
	\theta = \frac{ \gamma_{1} \sigma_{X}^{2} }{\gamma_{1}^{2} \sigma_{X}^{2} + \gamma_{2}^{2} \sigma_{h}^{2}    }
\end{equation}
Finally, it is immediate to get that conditional variance is: 
\begin{eqnarray}
	Var ( \tilde{S} \vert p ) & = & Var(\tilde{S}) - \frac{ Cov( \tilde{S}, p)^{2}  }{Var(p)}  \label{condvar1} \\
	& = & \sigma_{X}^{2} + \sigma_{h}^{2} - \frac{ \gamma_{1}^{2} \sigma_{X}^{4} }{\gamma_{1}^{2} \sigma_{X}^{2} + \gamma_{2}^{2} \sigma_{h}^{2}    } \label{condvar2}
\end{eqnarray}
The demand of non-informed agents (\ref{dema}) is now given by: 
\begin{equation}\label{dema2b}
	Y  =   \frac{ (\theta - 1)p   }{\alpha Var (\tilde{S} \vert p )   }  
\end{equation}
We are now in the position to state the following Theorem: 
\begin{theorem} \label{th2}
	The equilibrium price level for a Rational Expectations Equilibrium is: 
	\begin{equation}\label{pREE}
		p = \delta_{1} E(\tilde{S} \vert X) + \delta_{2} \tilde{h} 
	\end{equation}
where $ \delta_{1} $ and $ \delta_{2} $, are defined as follows: 
\begin{eqnarray}
	\delta_{1} & = & \frac{  N Var ( \tilde{S} \vert p)  }{N Var ( \tilde{S} \vert p)  +  (1-\theta) M \sigma_{\epsilon}^{2}} \label{delta1} \\
	\delta_{2} & = & \frac{  \alpha Var ( \tilde{S} \vert p) \sigma_{\epsilon}^{2} Z  }{N Var ( \tilde{S} \vert p)  +  (1-\theta) M \sigma_{\epsilon}^{2}} \label{delta2}
\end{eqnarray}
\end{theorem}

\begin{proof}
  See Appendix \ref{appD}.
\end{proof}
Clearly, the rule (\ref{pREE}) should be equivalent to the guessed expression (\ref{p}).  The solution for $ \gamma_{1} $ and $ \gamma_{2} $ is obtained by setting $\delta_{1} = \gamma_{1}  $, $\delta_{2} = \gamma_{2}  $.  It is easy to see that both (\ref{delta1})-(\ref{delta2}) - through the definition of $ Var ( \tilde{S} \vert p) $ given in (\ref{condvar2}) - are nonlinear function of parameters. The solution to be found is given by $ \gamma_{1}^{*}  $, $ \gamma_{2}^{*}  $, which are a function of all other parameters of the model. However, it is not immediate to find a solution ready to be intuitively interpreted. 

Therefore, to get a deeper sense of these results, let us proceed in two steps. First, let us make a simplifying assumption, by setting $ \gamma_{2} = 0 $.  Secondly, we simplify our setting to focus on a fully revealing Rational Expectations equilibrium.  

Let us start with the first case, postponing the second to the following subsection. Assume now that $ \gamma_{2} = 0 $: in this case, we do not have any noise traders in the economy. Thus, after taking advantage of (\ref{condvar2}), by setting  $\delta_{1} = \gamma_{1}  $, we find, after rearrangement the following solution for $ \gamma_{1}^{*} $: 
\begin{equation}\label{gamma1star}
	\gamma_{1}^{*} = \frac{1}{ 1 + \frac{ M (1 - \theta) \sigma_{\epsilon}^{2} \sigma_{h}^{2} }{ N \sigma_{X}^{2} }   }
\end{equation}
It is immediate to check that if the number of informed traders tends to infinity, the coefficient $ \gamma_{1}^{*} $ tends to 1: the equilibrium price converges to $ E(\tilde{S} \vert X) $. 

A similar result obtains when we set $ \gamma_{1} = 0 $: in this case, if  $ N \rightarrow \infty $, then $ \gamma_{2}^{*}  \rightarrow 1 $ and the price level converges directly to the noise trader signal.  

\textbf{In both cases, by increasing the number of market participants, makes prices more informative, making private information to become public, also by neutralizing the noise trading activity.}

\subsection{Fully Revealing Rational Expectations Equilibrium}

Let us assume that noise demand is non-stochastic, so that $ \tilde{h} = h $.  Recall the na\"{i}ve equilibrium price given by equation(\ref{pricena}), here recalled: 

	\begin{equation}\label{pricena2}
	p = \theta_{1} E (\tilde{S} \vert \tilde{X}) + (1 - \theta_{1}  ) E (\tilde{S}) + \theta_{2} h
\end{equation}

with $ \theta_{1} $ and $ \theta_{2} $ defined by (\ref{theta1}) and (\ref{theta2}), respectively.  

From equation (\ref{pricena2}), the only unknown is given by the signal $ \tilde{X}  $, which belongs to the information set of informed traders' only. Therefore, non-informed traders can learn about the signal by considering the actual price $ p $: in this way, they can use the price $ p $ to update both expected value and variance of future security's price $ \tilde{S} $ conditional to the observed market price $ p $. Clearly, this situation is close to a context where non-informed traders are conditioning on the signal $ \tilde{X} $. 

In this case the key results are condensed in the Theorem \ref{th3}. 
\begin{theorem} \label{th3}
	The equilibrium price for the Fully Revealing Rational Expectations Equilibrium (REE) is given by: 
	\begin{equation}\label{pfre}
		p = E (\tilde{S} \vert \tilde{X}  ) - \frac{ \alpha Z Var(\tilde{S} \vert \tilde{X}   )  h  }{N + M}
	\end{equation}
	where $ E (\tilde{S} \vert \tilde{X}  )  = X $, $ Var(\tilde{S} \vert \tilde{X}   ) = \sigma_{S} $. 
\end{theorem}
\begin{proof}
  See Appendix \ref{appE}.
\end{proof}
The result stated in Theorem \ref{th3} is absolutely important. \textbf{The equilibrium price $p$ differs from the expectation about future price given the signal, because of a correction due to volatility, risk aversion and noisy demand}: higher risk aversion $ \alpha $, higher is the volatility $ Var(\tilde{S} \vert \tilde{X}   ) $, or the demand of noisy traders $ h $, lower will be the price and, for any given value of the future's value $ \tilde{S} $ of the security's price, higher will be return: $ \tilde{R} = \tilde{S} - p $. 

Clearly, when $ N $ of $ M $ (or both) become absolutely large (i.e. when $ M, N \rightarrow \infty $), then the price level tend to be higher, converging to $ E (\tilde{S} \vert \tilde{X}  )  = X $. In this case, higher is the price, narrower will be the gap with respect to $ \tilde{S} $ and lower will be the return $ \tilde{R} = \tilde{S} - p $. 

\textbf{Therefore, when the number of traders (both informed and non-informed) increases indefinitely, the price level increases and the return gets lower}. The intuition behind this result goes as follows: \textbf{as the number of trader increases, the distortions derived from differential information tend to disappear, prices become fully informative and the compensation for the risk becomes smaller}. 

What are the impact on risk associated to the security? We need to compute the variance of price $ p $, whose expression is given as follows: 

\begin{eqnarray}
	Var(p) & = & Var(\tilde{S} \vert \tilde{X})  + \frac{ \alpha^{2} Z^{2} Var(\tilde{S} \vert \tilde{X}   )^{2}  h^{2}  }{(N + M)^{2} }  -   2 E (\tilde{S} \vert \tilde{X}  )  \frac{ \alpha Z Var(\tilde{S} \vert \tilde{X}   )  h  }{N + M}         =       \label{variapok1}  \\
	& = & Var(\tilde{S} \vert \tilde{X}) + \frac{ \alpha Z h Var(\tilde{S} \vert \tilde{X}   )    }{N + M} \left[ \frac{ \alpha Z h Var(\tilde{S} \vert \tilde{X}   )    }{N + M}  - 2   E (\tilde{S} \vert \tilde{X}  )        \right]     \label{variapok2}
\end{eqnarray}

The expression inside the square brackets in (\ref{variapok2}) is positive if: i) the degree of risk aversion $ \alpha $ is sufficiently high; ii) the number of noise traders $ Z $ is high; iii) $ Var(\tilde{S} \vert \tilde{X})$ is high; iv) $ N $, $ M $ are small: all circumstances easily verified.  

It is immediate to check from (\ref{variapok2}), that if $  M, N \rightarrow \infty $, $ Var(p) $ becomes smaller, and tend to the conditional variance $ Var(\tilde{S} \vert \tilde{X}) $. 

The informational efficiency index $ IE $ is defined as the reciprocal of the variance: 
\begin{equation}\label{ie}
	IE = \frac{1}{Var(p)}
\end{equation}
When $ Var(p) $ becomes smaller, $ IE $ increases. 

Summing up we found that, increasing the number of agents in a fully revealing REE, makes prices more informative and converging to privately owned information. At the same time, the volatility of the price becomes lower, as well as informational efficiency. The return is lower than in the case of a low number of informed or uninformed traders. \textbf{Moreover, as the number of agents becomes progressively larger, the degree of risk aversion becomes negligible.} 

By increasing the number of market participants, private information becomes publicly available, thanks to the assumption that non-informed traders can learn about the private signals, by monitoring current equilibrium market prices. 

\subsection{Concluding Remarks}

Building on insights from \cite{kyle89,rindi08,djrindi}, we have examined how varying the number of informed and uninformed participants influences equilibrium asset pricing. In both naïve scenarios without belief updates and more sophisticated frameworks where uninformed agents refine their assessments based on current prices, a key finding remains the same: increasing the number of participants amplifies the market's informative efficiency and lowers price volatility. Although larger markets tend to have lower returns--reflecting a reduced need to compensate traders for potential adverse price movements--they benefit from broader liquidity, more robust price discovery, and greater collective access to information.

By attracting a wide spectrum of users, RIV Coin \textbf{leverages these principles to establish a stable reserve in the Vault}. The increased participation directly mitigates volatility and enhances price accuracy, ensuring that RIV Coin functions reliably as a medium of exchange, a numeraire, and a store of value. In essence, the very openness of RIV Coin's ecosystem enables it to capture the equilibrium advantages outlined in the theoretical results, creating a more transparent, liquid, and resilient market for all participants.

\section{Solution}
RIV Coin is a cryptocurrency backed by a diversified set of reserves (the Vault) pledged in the exclusive interest of its users. The Vault is actively invested in a diversified CeFi/DeFi solution that does so under transparent, on-chain AND off-chain governance. 

The Vault is composed of two interconnected entities:

\begin{enumerate}
  \item A professionally managed, segregated, compartment of an alternative fund containing real-world assets, whose operational events and state transitions are aggregated using a zero-knowledge Proof of Integrity system, hereinafter called the ``Atlas'' (see Section \ref{sec:atlas})
  \item An on-chain reserve holding digital assets directly controlled by \textit{only} the protocol, hereinafter called the ``Aegis'' (see Section \ref{sec:aegis}) 
\end{enumerate}

Together, these elements provide a comprehensive, verifiable reserve that anchors the intrinsic value of RIV Coin.

\subsection{Atlas Ledger and Proof of Integrity}\label{sec:atlas}
\textit{Atlas} is a zk-Rollup specifically designed for investment fund operations, enabling transparent and verifiable fund accounting without revealing sensitive internal data.

\begin{figure}
  \centering
  \includegraphics[width=0.8\textwidth]{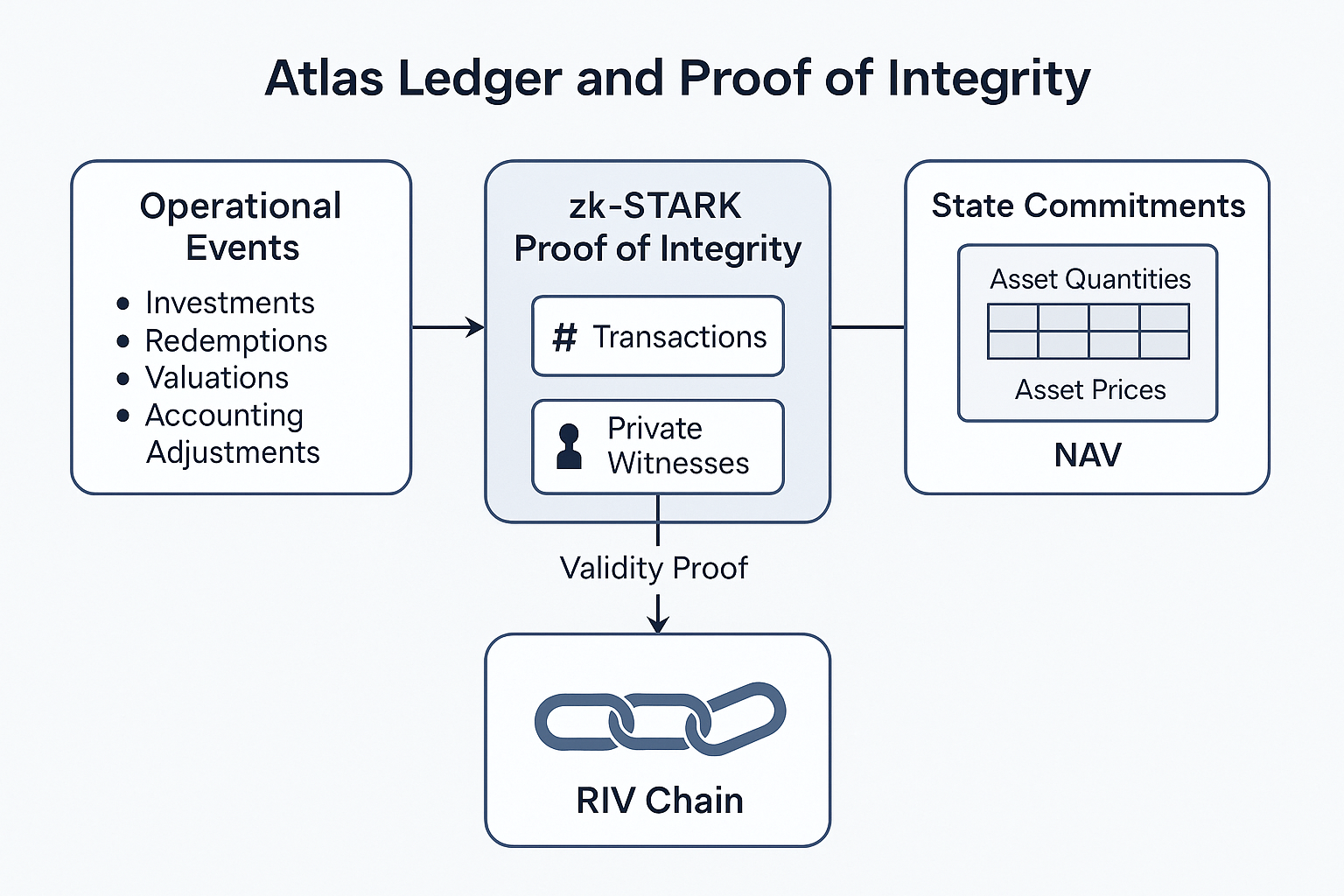}
  \caption{Atlas Proof of Integrity System}\label{fig:atlas}
\end{figure}

\begin{definition}[zk-Rollups]
Aggregators in zk-Rollups submit a succinct cryptographic proof (called a \emph{validity proof}) that asserts the correctness of a new published state root relative to the transaction data. This proof guarantees that the transaction execution was exact without requiring the verifier to recompute the operations. Zero-knowledge properties ensure that the necessary inputs remain confidential.\cite{rollups}
\end{definition}

Just as zk-Rollups aggregate off-chain transactions into verifiable proofs, \textit{Atlas} aggregates operational events and state transitions of a traditional asset portfolio—including investments, redemptions, valuations, and accounting adjustments—into cryptographic proofs of integrity.

\paragraph{State Definition.}
Each ledger \textbf{state} $S$ consists of:
\begin{itemize}
  \item A public homomorphic vector commitment to the full fund state,
    \begin{itemize}
      \item Instantiated as a Pedersen vector commitment $\mathcal{C}(S)$, satisfying
      \[
        \mathcal{C}(S + S') = \mathcal{C}(S) \cdot \mathcal{C}(S'),
      \]
      ensuring additive homomorphism.
      \item Internally includes:
      \begin{itemize}
        \item A vector of asset quantities,
        \item A vector of asset prices.
      \end{itemize}
    \end{itemize}
  \item A public Net Asset Value (NAV) field stored separately in cleartext after each block.
\end{itemize}

Asset quantities, prices, and NAV are represented as fixed-point integers over the prime field $\mathbb{F}$, scaled appropriately to prevent rounding errors and modular wraparound. The zk-STARK circuit enforces that, after sequentially applying the batch of transactions, the publicly stored NAV matches the dot product of the committed asset quantities and prices:
\[
\text{NAV} = \langle \text{assets}, \text{prices} \rangle.
\]
Within the circuit, range constraints ensure all values remain within safe bit-length bounds to avoid wrap-around.

\paragraph{System Design.}
\textit{Atlas} implements a Zero Knowledge Scalable Transparent ARgument of Knowledge (zk-STARK)\cite{starks} over the relation
\begin{equation}\label{eq:stark_rel}
R \subseteq \mathcal{X} \times \mathcal{W},
\end{equation}
where:
\begin{itemize}
  \item $\mathcal{X}$ is the set of public inputs,
  \item $\mathcal{W}$ is the set of private witnesses,
  \item $(x,w) \in R$ if and only if $w$ satisfies the correctness predicate for $x$.
\end{itemize}

In \textit{Atlas}:
\begin{itemize}
  \item $x = (S_0, S_n, h, \mathrm{NAV})$, where:
    \begin{itemize}
      \item $S_0$ is the homomorphic commitment to the initial asset and price vectors,
      \item $S_n$ is the homomorphic commitment to the final asset and price vectors,
      \item $h$ is the Merkle root of the transaction batch,
      \item $\mathrm{NAV}$ is the public Net Asset Value after processing the batch.
    \end{itemize}
  \item $w = (T, \mathrm{assets}, \mathrm{prices})$, where:
    \begin{itemize}
      \item $T$ is the ordered list of transactions.
      \item $\mathrm{assets}$ is the assets vector.
      \item $\mathrm{prices}$ is the prices vector.
    \end{itemize}
\end{itemize}

The verification condition is that there exists a valid witness $w$ such that:
\begin{align}
R(x, w) \iff\ 
& S_n = \mathcal{C}(\mathcal{A}(S_0, T)) \nonumber \\
& \land\ h = \texttt{MerkleRoot}(T) \nonumber \\
& \land\ \text{NAV} = \langle \text{assets}, \text{prices} \rangle
\end{align}
where:
\begin{itemize}
  \item $\mathcal{A}(S_0, T)$: Sequentially applies the transactions $T$ starting from $S_0$
  \item $\mathcal{C}(S)$: Computes the cryptographic commitment to a state $S$,
  \item $\texttt{MerkleRoot}(T)$: Computes the Merkle root over the batch of ordered transactions.
\end{itemize}

Only the Merkle root $h$ of transactions $T$ is made public; the full transaction list $T$ remains confidential unless individual transactions are selectively disclosed via Merkle proofs by the committee. All batches are padded to a fixed size (e.g., next power of two) with dummy transactions to avoid metadata leakage.

\paragraph{Transaction Types.}
Transactions belong to a predefined set:
\begin{align}
\mathcal{T} =\ 
& (\{\texttt{Deposit}\} \times \mathcal{U} \times \mathbb{F}) \nonumber \\
\sqcup\ 
& (\{\texttt{Withdrawal}\} \times \mathcal{U} \times \mathbb{F}) \nonumber \\
\sqcup\ 
& (\{\texttt{AssetPurchase}\} \times \mathbb{F} \times \mathbb{F} \times \mathbb{F}) \nonumber \\
\sqcup\ 
& (\{\texttt{AssetSale}\} \times \mathbb{F} \times \mathbb{F} \times \mathbb{F}) \nonumber \\
\sqcup\ 
& (\{\texttt{ManagementFee}\} \times \mathbb{F}) \nonumber \\
\sqcup\ 
& (\{\texttt{PerformanceFee}\} \times \mathbb{F}) \nonumber \\
\sqcup\ 
  & (\{\texttt{MarkToMarket}\} \times \mathbb{F}^n)
\end{align}

where $\mathcal{U}$ is the set of user addresses and $\mathbb{F}$ is the prime field underpinning the zk-STARK.
Each transaction is defined in Algorithm \ref{alg:structdefs}:

\begin{algorithm}[!h]
\caption{Transaction Definitions}\label{alg:structdefs}
\begin{algorithmic}[1]
\STATE \textbf{type} Deposit
\STATE \hspace{1em} Amount: Fiat amount deposited
\STATE \hspace{1em} User: User ID of the depositor
\STATE
\STATE \textbf{type} Withdrawal
\STATE \hspace{1em} Amount: Fiat amount withdrawn
\STATE \hspace{1em} User: User ID of the withdrawer
\STATE
\STATE \textbf{type} AssetPurchase
\STATE \hspace{1em} Index: Index of the asset purchased in the assets vector
\STATE \hspace{1em} Quantity: Amount of the asset purchased
\STATE \hspace{1em} Price: Effective unit price of the asset at purchase, net of fees.
\STATE
\STATE \textbf{type} AssetSale
\STATE \hspace{1em} Index: Index of the asset purchased in the assets vector
\STATE \hspace{1em} Quantity: Amount of the asset sold
\STATE \hspace{1em} Price: Effective unit price of the asset at sale, net of fees.
\STATE
\STATE \textbf{type} ManagementFee
\STATE \hspace{1em} FeeAmount: Management fee deducted
\STATE
\STATE \textbf{type} PerformanceFee
\STATE \hspace{1em} FeeAmount: Performance fee deducted
\STATE
\STATE \textbf{type} MarkToMarket
\STATE \hspace{1em} NewPrices: Vector of updated asset prices
\end{algorithmic}
\end{algorithm}

The \texttt{MarkToMarket} transaction updates the full vector of asset prices at once, where $n$ is the amount of assets, and by extension prices.

Within each block, only the final \texttt{MarkToMarket} transaction is used to update the prices vector for purposes of NAV computation. Intermediate \texttt{MarkToMarket} transactions, if present, are ignored.

\paragraph{Ledger Structure.}
Each batch of operational transactions is structured as a \textbf{block} containing:
\begin{itemize}
  \item A block number indicating the ledger sequence,
  \item The Merkle root $h$ of the ordered transactions included in the block,
  \item The old state commitment $S_0$ and the new state commitment $S_n$,
  \item The public NAV value after processing the block.
\end{itemize}

Each proof $\pi$ associated with a block is co-signed by fund administrators and posted on RIV Chain, where the dedicated \texttt{x/atlas} module verifies it against the relation $R$. State commitments are chained: the new state $S_n$ from block $k$ becomes the old state $S_0$ for block $k+1$, ensuring sequential integrity.

\paragraph{Security and Transparency.}
By combining zk-Rollup architecture with transaction Merkle batching and succinct validity proofs, \textit{Atlas} achieves:
\begin{itemize}
  \item \textbf{Scalability}: Compressing many transactions into a single succinct proof,
  \item \textbf{Privacy}: Keeping sensitive financial operations confidential while still verifiable, with transaction batch padding to prevent metadata leakage,
  \item \textbf{Transparency}: Providing public, tamper-proof assurance of fund accounting integrity,
  \item \textbf{Selective Auditability}: Allowing individual transactions to be selectively revealed via Merkle proofs without exposing unrelated data.
\end{itemize}

Fund units of the vehicle of investment are entrusted to a Swiss trust governed by a committee of trustees. To ensure coherent oversight, the trustees are the same entities comprising the DAO's committee (detailed further in Section~\ref{sec:dao}), thus aligning on-chain governance with off-chain fiduciary obligations.

At launch, Atlas will focus on investing in a diversified Luxembourg SICAV RAIF (Reserved Alternative Investment Fund) which diversifies into Fixed Income listed securities, thematic equity portfolios (listed stocks), and a diversified number of mutual/hedge/ETFs funds. This provides efficient, capital-light access to a wide range of assets.
As Atlas accumulates more capital, it will pursue further diversification by reinvesting into additional asset classes.

\subsection{RIV Chain Architecture}
\begin{figure}[!h]
  \centering
  \includegraphics[width=0.6\textwidth]{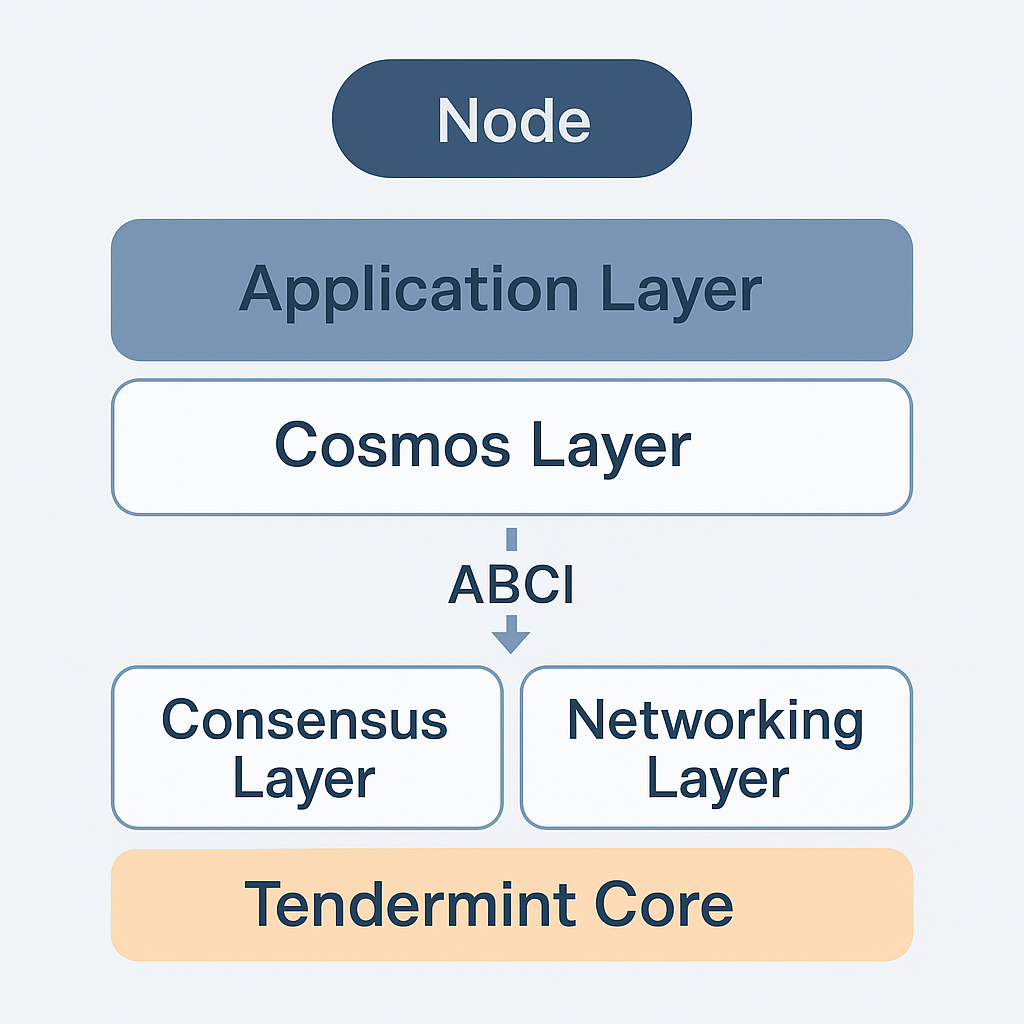}
  \caption{Cosmos-SDK Blockchain Architecture Overview}
  \label{fig:architecture_overview}
\end{figure}

RIV Chain is a Layer-1 sovereign blockchain implemented on top of the Cosmos-SDK. It uses a Proof-of-Stake consensus powered by CometBFT: a distributed, Byzantine fault-tolerant, deterministic state machine replication and consensus engine, which is the successor to Tendermint Core \cite{buchman2019latestgossipbftconsensus, kwon2014tendermint}. Figure \ref{fig:architecture_overview} illustrates the high-level architecture of a Cosmos-SDK-based blockchain, highlighting the separation between consensus and application logic. Communication between these layers is handled by the Application Blockchain Interface (ABCI), a protocol that allows CometBFT to remain agnostic to the specific application state machine.

\subsubsection{Modules}

\begin{figure}[!ht]
  \centering
  \includegraphics[width=0.4\textwidth]{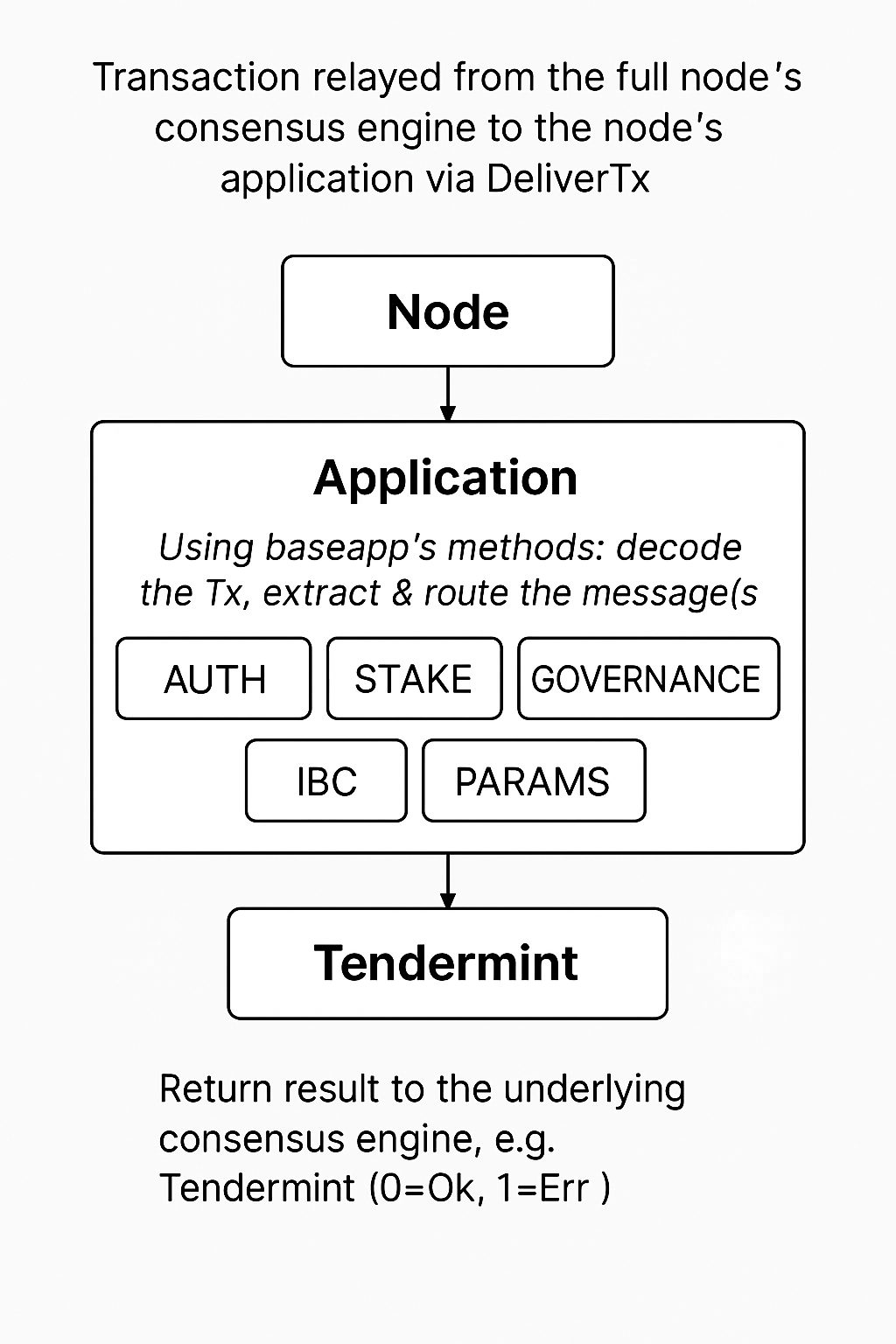}
  \caption{Cosmos-SDK Modules Message Processing}
  \label{fig:cosmos_modules}
\end{figure}

Cosmos SDK modules define the unique functional and stateful properties of each blockchain application.  
Modules can be viewed as independent state machines operating within the larger application-level state machine.  
They encapsulate critical elements of blockchain behavior, including storage layouts (state) and state transition functions (message handlers and transactions).  
Figure \ref{fig:cosmos_modules} illustrates how transactions are routed and handled by their respective modules.

For the purposes of this whitepaper, we will not be exhaustively covering all of the modules used by RIV Chain, such as the core modules and UX-related supplementary modules, but only the ones relevant to RIV Coin. 
Mainly, that includes the Inter-Blockchain Communication (IBC) modules, and the custom modules implemented by RIV Chain.

\paragraph{Inter-Blockchain Communication (IBC) Modules}

RIV Chain leverages the Inter-Blockchain Communication (IBC) protocol to interact with external blockchains and services.  
IBC allows RIV Chain to perform secure, trust-minimized communication across sovereign chains, enabling a wide range of cross-chain functionalities.
The primary IBC modules included are:

\begin{itemize}
  \item \texttt{x/core} -- Provides the underlying IBC protocol logic, including connection, channel, and packet lifecycle management.
  \item \texttt{x/apps} -- Implements application-layer protocols built on top of IBC, such as Interchain Accounts (ICA) and Interchain Queries (ICQ).
  \item \texttt{x/light-clients} -- Manages the verification of counterparty chain states via light clients (e.g., Tendermint-based light clients).
\end{itemize}

RIV Chain's use of IBC includes:
\begin{itemize}
  \item \textbf{Interchain Accounts} -- Allowing RIV Chain to control accounts on remote chains like Osmosis, enabling seamless liquidity management and DeFi interactions without bridging.
  \item \textbf{Osmosis Liquidity Pools} -- Directly owning and managing liquidity positions in Osmosis pools through Interchain Accounts, enabling native liquidity provision and yield strategies.
  \item \textbf{Band Protocol Oracles} -- Accessing off-chain price feeds via Band Protocol's IBC-enabled oracles to obtain reliable real-time asset pricing for portfolio valuation and risk management.
\end{itemize}

\paragraph{Custom Modules}

Last, but not least, the custom modules purpose built for RIV Chain and RIV Coin:
\begin{itemize}
  \item \texttt{x/atlas}
  \item \texttt{x/aegis}
  \item \texttt{x/emissions}
  \item \texttt{x/escrow}
\end{itemize}

\texttt{x/atlas} was explained in Section \ref{sec:atlas}, the remaining are explained in the following sections.

\subsubsection{Aegis}\label{sec:aegis}
\textit{Aegis} refers to a special account on the blockchain that is not governed by private keys.
Instead, the assets held within Aegis can only be moved or utilized through operations explicitly defined by the module itself.
This design enables the protocol to directly own and manage digital assets without relying on individual custodians.
Assets under Aegis may reside natively on the RIV Chain, or, through the Interchain Accounts (ICA) standard, on other IBC-enabled blockchains.
At the time of writing, Aegis supports two authorized operations:
\begin{enumerate}
    \item Transfers to \textit{Atlas}
    \item Protocol-Owned Liquidity (POL) operations (see Section \ref{sec:pol})
\end{enumerate}

For an explanation of how assets are deposited into Aegis, refer to Section \ref{sec:mintingflow}.

Aegis lays the foundation for broader autonomous DAO-controlled execution management, with additional capabilities currently under research and planned for future implementation.

\subsubsection{Escrow Mechanism}\label{sec:escrow}

\textit{Escrow} facilitates secure, auditable transfers of digital assets from \textit{Aegis} to the fund's custodied wallet—under the control of the regulated custodian bank.

The process begins with a \texttt{MsgInitiateEscrowTransfer}, where \textit{Aegis} moves specific assets into a protocol-controlled escrow account. This account grants the DAO committee multisig-based, least-privilege authorization to:

\begin{enumerate}[label=\alph*)]
  \item Execute transfers only to whitelisted addresses (i.e., the fund's custodied wallet).
  \item Authorize Private Key Control Procedure (PKCP) transactions required for KYC compliance.
\end{enumerate}

Once assets are released to the custodied wallet, it counts as a deposit or subscription by the DAO committee and the fund units issued are placed in the trustee's custody.
They can be off-ramped to fiat or traded by the invested fund, and are included in ongoing audits managed through \textit{Atlas}.

\begin{figure}[!h]
  \centering
  \includegraphics[width=0.7\textwidth]{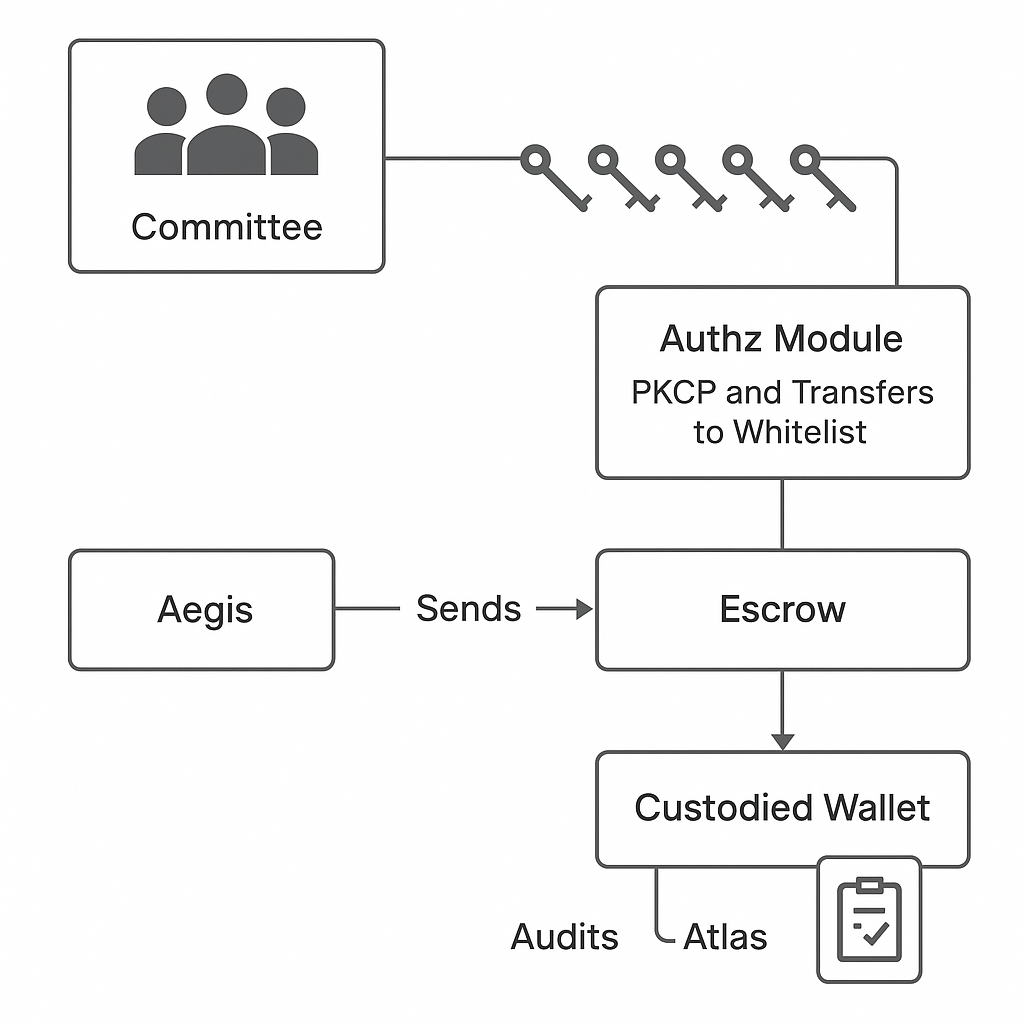}
  \caption{Escrow Mechanism Flow}
  \label{fig:escrow}
\end{figure}

\section{Tokenomics}
\label{sec:tokenomics}

In this section, we describe the Tokenomics of the RIV Coin ecosystem, covering the structure of governance, liquidity management, staking rewards, minting mechanisms, and the emissions system.

\subsection{Decentralized Autonomous Organization (DAO)}
\label{sec:dao}

Governance of the RIV ecosystem is decentralized and coordinated by a DAO structure. Decision-making authority is vested in a governance committee operating through a multisignature wallet (multisig). The committee is responsible for adjusting protocol parameters, enacting policy changes, and stewarding the health of the system, all while adhering to principles of transparency and democratic participation.

\subsubsection{Protocol-Owned Liquidity (POL)}
\label{sec:pol}

Protocol-Owned Liquidity (POL) refers to liquidity positions that are directly controlled by the DAO, rather than by individual liquidity providers. POL ensures that the protocol always maintains deep, reliable liquidity for RIV Coin markets, reducing slippage and improving price stability. Governance, via the committee's multisig, controls and allocates POL deployments across different pools and venues. The liquidity positions, however, are owned by \textit{Aegis}, via the Interchain Accounts standard, where the protocol controls accounts on other blockchains and DEXes, and receives the swap fees rewards.

\subsubsection{Staking}

Staking rewards are derived from a fraction of the net return generated by the Vault's treasury assets. The reward distribution occurs on a per-epoch basis, where the epoch length and reward fraction are configurable governance parameters. Adjustments to staking parameters are authorized through governance proposals ratified by the committee's multisig. This ensures that staking remains flexible to changing market conditions while protecting long-term sustainability.

\subsection{Minting Flows}

The circulating supply of RIV Coin can change through three mechanisms:

\begin{enumerate}
    \item \textbf{Inflationary Minting}: Minting that occurs independently of any new reserve deposits in the Vault.
    \item \textbf{Non-Inflationary Minting}: Minting backed by verifiable reserve deposits into the Vault.
    \item \textbf{Burning}: Permanent destruction of tokens, reducing circulating supply.
\end{enumerate}

The staking module manages inflationary minting, which distributes minted RIV Coin as rewards to liquidity providers. The committee's multisig retains discretionary authority to initiate inflationary minting within strict constraints, including a \textbf{maximum cap of 100\% of the total circulating supply} enforced at the DAO level.

Non-inflationary minting follows a secured and verifiable flow (detailed in Section~\ref{sec:mintingflow}), wherein minted RIV Coin corresponds to reserves pledged to the Vault. The result is a quasi-full hedge, where minted tokens are effectively backed always by the value of assets under custody.

The robustness of this reserve-backed design ensures that systemic collapse would require catastrophic external events (e.g. Lehman default, 2022 crisis) or extraordinary fraud (e.g., Madoff, FTX scandals), both of which are mitigated by segregated accounts, multi-layered proofs, and conservative custody policies. Notably, assets within the Vault cannot be rehypothecated or pledged as collateral.

RIV Coin rests upon the three pillars of \textbf{MPH}:

\begin{itemize}
    \item \textbf{Meritocracy}: Users who participate in pooling and staking are proportionally rewarded.
    \item \textbf{Pragmatism}: Governance decisions and code implementations prioritize user protection and adaptability.
    \item \textbf{Honesty}: Transparent, predictable rule sets ensure fair play with no hidden mechanisms.
\end{itemize}

The combination of advanced proof algorithms, transparent fund segregation, and strict compliance frameworks ensures that price discovery remains free, organic, and legally sound.

\subsubsection{The Emissions Manager and Non-Inflationary Minting}
\label{sec:mintingflow}

The Emissions Manager (EM) is a core component of the DAO responsible for managing controlled expansions of RIV Coin supply into the market through non-inflationary means.

Initially, during the \textbf{Presale Phase}, fiat capital is deposited into the Vault, allowing RIV Coin to be minted at a fixed 1:1 price against reserves. Presale proceeds are used to continue minting new batches, with a portion allocated to bootstrap liquidity pools. After the presale, the protocol transitions into the \textbf{Open Market Phase}.

In the Open Market Phase, a Liquidity Pool (LP) is established where RIV Coin acts as the quote asset. Users can freely trade into and out of RIV Coin and participate as liquidity providers.

The Emissions Manager operates during this phase by monitoring market conditions. Every 3 epochs, it evaluates the \textit{premium} (market price / backing price) of RIV Coin and if the premium exceeds a configurable minimum premium threshold, the Emissions Manager mints and sells a computed quantity of RIV Coin on the open decentralized exchange in exchange for fresh reserves. This issuance amount is governed by the following equation:

\begin{equation}
  r_t = \sum_{i=0}^{t-1} r_i \times b \times \frac{p + 1}{P_{\text{min}} + 1}
\end{equation}

Where:

\begin{itemize}
    \item $r_t$ = Amount of RIV Coin emitted at epoch $t$
    \item $b$ = Base emissions rate
    \item $p$ = Current premium (market price / backing price)
    \item $P_{\text{min}}$ = Minimum premium threshold
\end{itemize}

\begin{figure}[!h]
  \centering
  \includegraphics[scale=0.7]{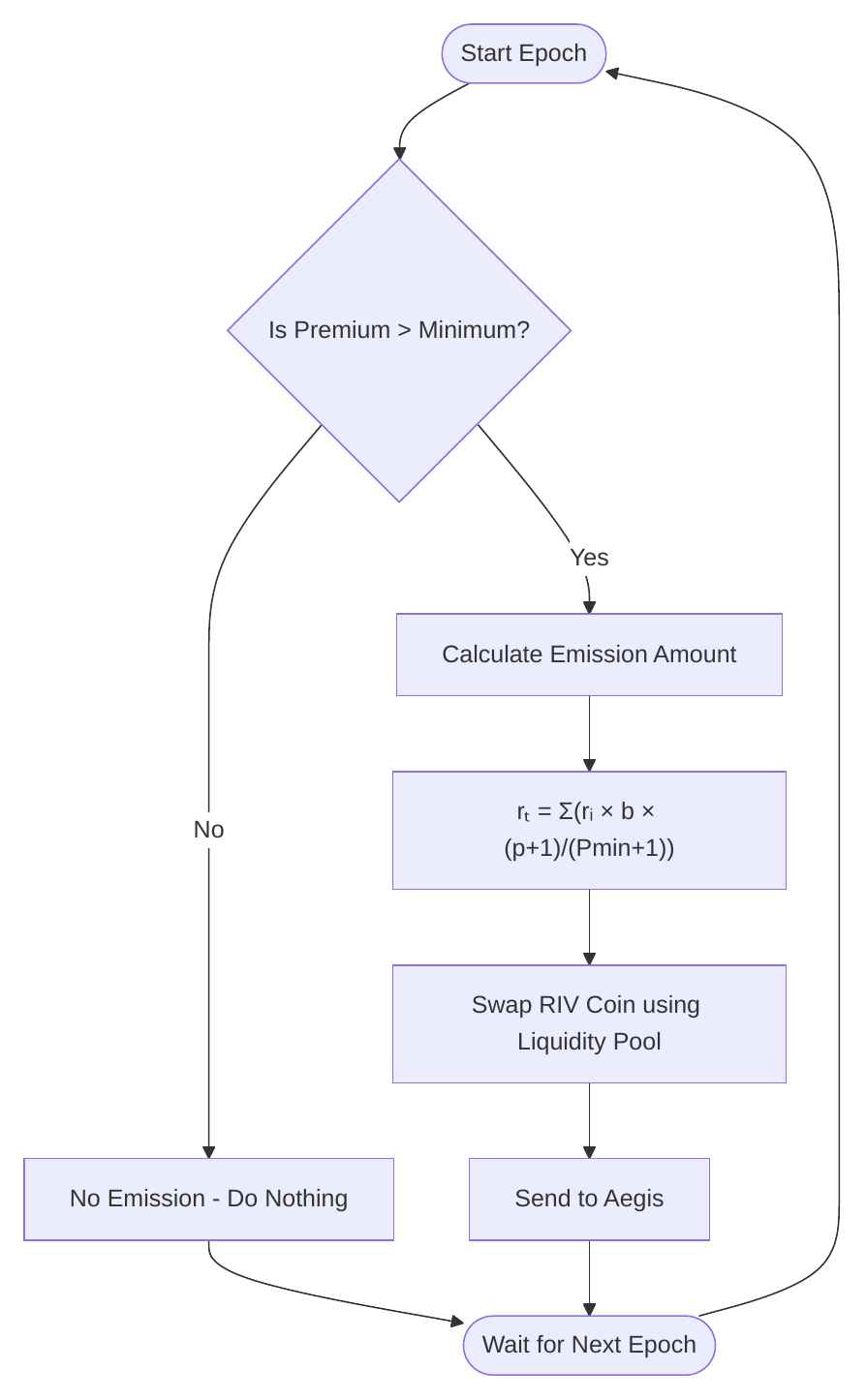} 
  \caption{Minting flow}
  \label{fig:minting}
\end{figure}

The flow of the Emissions Manager system is illustrated in Figure~\ref{fig:minting}.

Due to the premium requirement, this process is slightly deflationary, as the proceeds of the sale are greater than the intrinsic value of the minted coins. This results in an arbitrage that is completely in favor of the users, increasing the intrinsic value of their RIV Coin. Also as a consequence, the market price is closer to the intrinsic value, due to both the price impact of the sale and the deflationary nature of the mint.

To prevent a scenario where the founders control a disproportionate share of the token supply from the outset, the allocation of rewards to the Founder's Wallet is designed to be gradual and proportional to ongoing minting activity. A fixed and reasonable portion, capped at 5\% of the total token supply, will be allocated to a dedicated account known as the Founder's Wallet. This Wallet will not receive its full allocation at genesis. Instead, it will be gradually filled over time, in sync with user minting activity. 
For each user mint, the system will mint a slightly higher amount of tokens, calculated as $mx$, where $x = \frac{5\%}{1-5\%}$, and $m$ is the base mint amount. The additional portion represents the premium that goes directly to the Founder's Wallet. The more users mint, the faster the Founder's Wallet reaches its capped allocation, ensuring transparency, fairness and alignment between the founder and the community.  

\subsubsection{Natural Arbitrage Mechanisms}
RIV Coin's fully backed design positions it as the quote asset in various liquidity pools, effectively anchoring the relative pricing of other tokens.
Rather than depending solely on speculative market forces, RIV Coin's worth stems from the diversified reserves that underlie it.
Consequently, whenever a token's price--expressed in RIV Coin--deviates from that token’s price on other markets, arbitrageurs quickly capitalize on the discrepancy, driving the RIV Coin-denominated price back toward its intrinsic reserve value.
This interplay ensures that RIV Coin remains a stable reference, fostering reliable trading pairs and reinforcing its role as a dependable exchange medium.

\subsection{Governance}\label{sec:governance}

The DAO operates under a mixed governance model where authority is divided between decentralized tokenholders and an executive Committee operating via a multisignature (multi-sig) wallet. Certain critical operational functions fall under the purview of the Committee for efficiency and risk mitigation, while broader strategic development is directed by tokenholder votes.

All RIV Coin possess equal voting rights, with $n$ RIV equating to $n/S$ of the total voting power, where $S$ is the total circulating supply.

\subsection{Tokenholder Voting Authority}

The following matters are decided through decentralized voting by all RIV Coin holders:
\begin{itemize}
    \item \textbf{Feature Requests}: Tokenholders may propose and vote on new feature implementations, expansions, and initiatives beyond the protocol's existing scope.
    \item \textbf{Sale of Last Resort}: Approval of liquidation of the Vault's assets and token redemption at intrinsic value during systemic crises.
\end{itemize}

Voting is conducted on a one-token, one-vote basis, with quorum and approval thresholds established in the DAO's governance documents.

\paragraph{Committee Authority}

The Committee serves as the executive arm of the DAO and operates via a multisignature wallet requiring threshold approval for actions. The following matters fall under the Committee's purview:
\begin{itemize}
    \item \textbf{Minting Authorization}: Approval of inflationary and non-inflationary minting operations within protocol-defined frameworks.
    \item \textbf{Protocol-Owned Liquidity (POL) Management}: Allocation and reallocation of liquidity deployments.
    \item \textbf{Emergency Response}: Execution of protective actions such as trading halts and liquidity withdrawal freezes during crises.
    \item \textbf{Dynamic Parameter Management}: Adjustment of staking rates, emission schedules, and premium thresholds.
    \item \textbf{Upgrades and Migrations}: Management and authorization of protocol upgrades, infrastructure migrations, and architectural changes.
    \item \textbf{Fund Management Oversight}: Appointment and supervision of the external asset managers responsible for the Vault's investments.
\end{itemize}

Committee members act as fiduciaries of the protocol and are subject to disclosure obligations and periodic accountability mechanisms.

\paragraph{Reserves Allocation}

Portfolio allocation of the Vault's reserves falls under the purview of the DAO Committee. As detailed in Section~\ref{sec:atlas}, initial investments focus on centralized, regulated vehicles with a track record of performance. Investment strategies aim for a prudent balance between risk and return, exclusively prioritizing the interests of RIV Coin holders.

\paragraph{Configuration Parameters}

The following dynamic parameters are managed by the Committee and may be adjusted periodically based on market conditions and protocol performance:

\begin{itemize}
    \item \textbf{Staking Rate}: The fraction of the Vault's net yield distributed to staking participants per epoch.
    \item \textbf{Base Emission Rate}: The fundamental rate at which new RIV Coin is emitted via bond markets, subject to premium conditions.
    \item \textbf{Minimum Premium}: The minimum premium (market price relative to backing value) required to trigger new emissions into the market.
\end{itemize}

These parameters are critical to balancing the protocol's growth dynamics, ensuring stability, aligning incentives, and maintaining reserve-backed value over time.

\paragraph{Migration Mechanism}

The authority to manage migrations and upgrades falls under the purview of the Committee. If necessary, the Committee may authorize migration to new contracts or architectures to enhance protocol functionality, security, or scalability.

\paragraph{Sale of Last Resort}

The sale of last resort is a protective mechanism allowing tokenholders to exit the system at intrinsic value during extreme adverse events.

\begin{itemize}
    \item \textbf{Initiation}: The DAO Committee may propose activating the sale of last resort.
    \item \textbf{Approval}: Requires approval through a tokenholder vote.
    \item \textbf{Execution}: Upon approval, the Vault's assets are liquidated, and proceeds are used to repurchase and burn RIV Coins at the current NAV intrinsic price.
\end{itemize}

This mechanism ensures that tokenholders retain the right to redeem real value even under distressed market conditions.

\subsection{Performance-Based Reserve Management}\label{sec:fundfees}

The fund that \textit{Atlas} tracks, which constitutes the centralized component of the Vault's reserves, is subject to standard asset management fees designed to sustain professional oversight, maintain operational excellence, and ensure long-term alignment with tokenholder interests.

The fee structure is composed of two elements:

\begin{itemize}
    \item \textbf{Performance Fees}: Fees applied to positive net performance of the fund, designed to align the incentives of the asset managers with the success of the Vault.
    \begin{itemize}
        \item 0--9\% NAV increase (monthly): 10\% performance fee
        \item 9--20\% NAV increase (monthly): 15\% performance fee
        \item 20\%+ NAV increase (monthly): 25\% performance fee
    \end{itemize}
    Performance fees are calculated on a monthly basis relative to the Net Asset Value (NAV) growth and are deducted from the gross returns before they are reflected in the Vault's proofs and reporting.

    \item \textbf{Management Fee}: A fixed annualized fee of 2\%, charged pro rata and deducted on a monthly basis from the fund's total assets under management (AUM).
\end{itemize}

These fees are industry-standard for actively managed portfolios with diversified and sophisticated investment strategies. They compensate professional asset managers for research, portfolio construction, risk management, legal compliance, auditing, and custodial arrangements critical to the sustainable growth of the reserves.

Importantly, all fees are already incorporated into the Net Asset Value (NAV) calculations reflected in Atlas proofs. Users interacting with RIV Coin or referencing the Vault's performance are always observing figures \textit{net of fees}, ensuring full transparency and eliminating hidden charges.

The fee structure is designed with the following principles:
\begin{itemize}
    \item \textbf{Alignment of Incentives}: Managers are rewarded only when net positive returns are achieved, ensuring their goals align with those of tokenholders.
    \item \textbf{Verifiability}: All fee deductions are reflected on-chain through Atlas' Proof of Integrity and are auditable.
    \item \textbf{Fairness}: No performance fee is charged in periods of negative returns, and fees scale progressively with fund performance to avoid excessive extraction.
\end{itemize}

Through this structure, the fund achieves a balance between incentivizing professional active management and preserving maximum value accrual for RIV Coin holders.

\section{Conclusion} 

In the real world, both in the old and the new economy, no substantial progress has been achieved regarding the biases caused by wealth inequality.
The problem of inequality has been perceived mostly as a political matter dividing the electorates between left and rightwing.
The reality is that aggressive taxation can only produce conflicts.
Direct taxes are often too high in general to be raised further and indirect taxes usually increase income inequality. 

We have explored the problem from a different point of view.
Let's bear in mind this analogy: \textbf{as well as we have to be creators and not competitors, we found the way to create and add wealth instead of causing its dispersion}.

No doubt that such a cryptocurrency will increase the capacity of the wealthy people, but the marginal utility is bigger for Retail Users who will benefit, with the placet of Wealthier Users, from the bigger amounts invested by these ones. The last ones look for stability, the first ones would like not to be frauded participating at a fair game.

The Blockchain Revolution will enable a fairer world \cite{bcstatistics} where everyone can profit. We believe that anyone has the right to live a life at its fullest potential: by creating RIVCoin, we want to allow you to achieve this objective by, to some extend simply ``leader following''. 

\printbibliography

\pagebreak
\begin{appendices}
	\section{Projection Theorem}{\label{appA}}
	%\section{Projection Theorem}
	Given two random variables $ \tilde{X} $, $ \tilde{Y} $, both distributed normally, conditional mean and
	variance of $ \tilde{X} \vert \tilde{Y}  $ can be recovered according to standard projection Theorem for Normal
	Distributions, such that:
	\begin{equation}\label{a1}
		E( \tilde{X} \vert \tilde{Y}  ) = E(\tilde{X} ) +   \frac{ Cov( \tilde{X} \vert \tilde{Y}   ) }{Var(\tilde{Y} )}    (  \tilde{Y}  - E(\tilde{Y}  )    )
	\end{equation}
	 Q.E.D. \qedsymbol
	
  \section{Proof of Proposition \ref{pro1}}{\label{appB}}
	
	Consider the following prior distribution $ f_{pr}(\tilde{S})  \sim N ( S, \sigma_{S}^{2})$, whose kernel is given as: \\ 
	\begin{equation}\label{kernpr}
		f_{pr}(\tilde{S}) \sim \exp\left\{ - \frac{1}{2 \sigma_{S}^{2} }     \left(   \tilde{S} - S \right)^{2}  \right\}
	\end{equation} \\
	The conditional likelihood $f_{L}(  \tilde{X} \vert  \tilde{S}) \sim N ( \tilde{S}, \sigma_{X}^{2}) $ is: 
	\\ 
	\begin{equation}\label{kernlike}
		f_{L}(  \tilde{X} \vert  \tilde{S}) \sim \exp\left\{ - \frac{1}{2 \sigma_{X}^{2} }     \left(   \tilde{X} - \tilde{S} \right)^{2}  \right\}
	\end{equation}
	\\
	Therefore the posterior: 
	\begin{eqnarray}
		f_{po}(  \tilde{S} \vert  \tilde{X}) & = & \exp \left\{    - \frac{1}{2 \sigma_{S}^{2} } (  \tilde{S}^{2} + S^{2} - 2 S \tilde{S}   )   - \frac{( \tilde{X}_{2} + \tilde{S}^{2}  - 2 \tilde{S} \tilde{X}    )}{2 \sigma_{X}^{2}  }         \right\}      =   \label{post1} \\
		& = & \exp \left\{   - \frac{\tilde{S}^{2}}{2 }   \left( \frac{1 }{\sigma_{S}^{2}}  + \frac{1}{\sigma_{X}^{2}}    \right)   + \tilde{S} \left( \frac{S}{ \sigma_{S}^{2} } + \frac{\tilde{X}}{ \sigma_{X}^{2} }   \right)   -  \frac{1}{2} \left(   \frac{S^{2}}{ \sigma_{S}^{2} }  +  \frac{\tilde{X}^{2}}{ \sigma_{X}^{2} }   \right)     \right\} \label{post2}
	\end{eqnarray} \\ 
	Focus now on the term in curly brackets of equation (\ref{post2}) which can be rewritten as: 
\\
	\begin{eqnarray}
		- \frac{1}{2} \left( \frac{1 }{\sigma_{S}^{2}}  + \frac{1}{\sigma_{X}^{2}}    \right) &  =  & -  \frac{1}{2 \sigma_{n}^{2}} \left(  \tilde{S}^{2} - 2 \tilde{S} \mu_{n} + \mu_{n}^{2}  \right)   \label{espo} \\
		& = & -  \frac{1}{2 \sigma_{n}^{2}} \left(  \tilde{S}  - \mu_{n}    \right)^{2} \label{espo2}
	\end{eqnarray} \\
	where: \\ 
	\begin{equation}\label{defi}
		\frac{1}{\sigma_{n}^{2}} = \frac{1 }{\sigma_{S}^{2}}  + \frac{1}{\sigma_{X}^{2}} 
	\end{equation}
	\\
	or, given the precision: \\ 
	\begin{equation}\label{defi2}
		\lambda_{S} =  \frac{1 }{\sigma_{S}^{2}}    \qquad  \lambda_{X} = \frac{1}{\sigma_{X}^{2}}
	\end{equation} \\ 
	implying:
	\begin{equation}\label{defi3}
		\lambda_{n} = \frac{1}{\sigma_{n}^{2}}  = \lambda_{S} + \lambda_{X}
	\end{equation}
	\\
	and, from (\ref{espo2}): 
	\\
	\begin{eqnarray}
		\mu_{n} & =  & \frac{\lambda_{S} S + \lambda_{X} \tilde{X} }{\lambda_{S} + \lambda_{X}} = \label{defi4} \\
		& = & \frac{\lambda_{S}}{\lambda_{S} + \lambda_{X}} S + \frac{\lambda_{X}}{\lambda_{S} + \lambda_{X}} \tilde{X}  \label{defi5} = \\
		& = & \mu_{1} S + (1 - \mu_{1}) \tilde{X} \label{defi6}
	\end{eqnarray}
	with: 
	\begin{equation}\label{defi7}
		\mu_{1} = \frac{\lambda_{S}}{\lambda_{S} + \lambda_{X}}
	\end{equation}
	as stated. \qedsymbol
	
	\pagebreak
  \section{Proof of Theorem \ref{th1}}{\label{appC}}
	Consider Market Equilibrium condition in (\ref{equili}) and substitute out the demand for risky security for non-informed and informed traders, given, respectively by (\ref{Y}) and (\ref{Yi}).  After some steps, we have: 
	\\
	\begin{eqnarray} 
		\frac{N  E(\tilde{S} \vert \tilde{X}) }{\alpha Var(\tilde{S} \vert \tilde{X}  )  } + \frac{M}{\alpha Var(\tilde{S} )} E(\tilde{S} ) + Z \tilde{h} & = & \left[  \frac{N}{\alpha Var(\tilde{S} \vert \tilde{X}  )  }  +  \frac{M}{\alpha Var(\tilde{S})  } \right] p    \label{teo1a}  \\
		& = & \left[  \frac{N Var(\tilde{S} \vert \tilde{X}) + M Var(\tilde{S})    }{\alpha Var(\tilde{S} \vert \tilde{X}) Var(\tilde{S})   }   \right] p  \label{teo1b}
	\end{eqnarray}
	\\
	Therefore, solving for $ p $ we obtain, after simplifying: 
	\\
	\begin{eqnarray}
		p  =  \left[  \frac{ N Var(\tilde{S})  }{N Var(\tilde{S} \vert \tilde{X}) + M Var(\tilde{S}) } \right] E(\tilde{S} \vert \tilde{X})  
		 & + & \left[  \frac{ M Var(\tilde{S} \vert \tilde{X}  )   }{N Var(\tilde{S} \vert \tilde{X}) + M Var(\tilde{S}) } \right] E(\tilde{S})  +  \nonumber \\		 
		 & + & \left[  \frac{ \alpha Var(\tilde{S} \vert \tilde{X}  ) Var(\tilde{S})   }{N Var(\tilde{S} \vert \tilde{X}) + M Var(\tilde{S}) } \right] Z \tilde{h}  \label{teo1c}
	\end{eqnarray}
	which proves the stated result. \qedsymbol

	\section{Proof of Theorem \ref{th2}}{\label{appD}}

	After substitution of (\ref{dema2}) and (\ref{dema2b}) into market equilibrium condition (\ref{equili}), we have: 
	\\
	\begin{equation}\label{equite2}
			N  \left(   \frac{  E(\tilde{S} \vert \tilde{X}  )  -   p   }{\alpha Var(\tilde{S} \vert \tilde{X}  )   } \right) + \frac{ M (\theta - 1 ) p   }{\alpha Var(\tilde{S} \vert p  )   }  + Z \tilde{h} =0
	\end{equation}
\\
	Recall that $ Var(\tilde{S} \vert \tilde{X}  )  =  \sigma_{\epsilon}^{2} $.  After rearrangement, we find: 
	\\
	\begin{equation}\label{equite2b}
		p \left[  \frac{N}{ \alpha \sigma_{\epsilon}^{2}  }  +  \frac{ M (\theta - 1 )   }{\alpha Var(\tilde{S} \vert p  ) }   \right] = N     \frac{  E(\tilde{S} \vert \tilde{X}  )    }{\alpha Var(\tilde{S} \vert \tilde{X}  )   } + Z \tilde{h}		
	\end{equation}
	\\
	\noindent which is equivalent to: 
	\\
	\begin{equation}\label{equite2c}
		p \left[    \frac{N Var(\tilde{S} \vert p  ) + (1 - \theta) \sigma_{\epsilon}^{2} M   }{ \alpha \sigma_{\epsilon}^{2} Var(\tilde{S} \vert p  ) }       \right] = N     \frac{  E(\tilde{S} \vert \tilde{X}  )    }{\alpha Var(\tilde{S} \vert \tilde{X}  )   } + Z \tilde{h}
	\end{equation}

\noindent Therefore: \\ 

\begin{equation}\label{equite2d}
	p = \left[  \frac{  \alpha \sigma_{\epsilon}^{2} Var(\tilde{S} \vert p  )    }{N Var(\tilde{S} \vert p  ) + (1 - \theta) \sigma_{\epsilon}^{2} M}         \right]     \frac{  E(\tilde{S} \vert \tilde{X}  )    }{\alpha \sigma_{\epsilon}^{2}    } N + \frac{  \alpha Z \sigma_{\epsilon}^{2} Var(\tilde{S} \vert p  )    }{N Var(\tilde{S} \vert p  ) + (1 - \theta) \sigma_{\epsilon}^{2} M}   \tilde{h}
\end{equation}

\noindent which after simplifying, proves the stated result.  \qedsymbol

	\section{Proof of Theorem \ref{th3}}{\label{appE}}
	Recall Market equilibrium condition: 
	\\
	\begin{equation}\label{mk1}
		N Y_{I} + M Y + Z h = 0 
	\end{equation}
\\ 
In this case, the demand of informed and non-informed agents are the same: 
\\ 
\begin{equation}\label{demaiapp}
	Y_{I} = Y =    \frac{  E(\tilde{S} \vert \tilde{X}  )  -   p   }{\alpha Var(\tilde{S} \vert \tilde{X}  )   }	
\end{equation}
\\
	Therefore, by substituting out the expression for the demand of informed agents $Y_{I}$ and non-informed, given by (\ref{demaiapp}), we get:
	\\
	\begin{equation}\label{equi}
		N  \left(   \frac{  E(\tilde{S} \vert \tilde{X}  )  -   p   }{\alpha Var(\tilde{S} \vert \tilde{X}  )   } \right) + M   \left( \frac{  E(\tilde{S} \vert \tilde{X}  )  -   p   }{\alpha Var(\tilde{S} \vert \tilde{X}  )   } \right) + Z h =0
	\end{equation}
\\
Which, after rearrangement, implies directly the results. Q.E.D. \qedsymbol 

\end{appendices} 

\end{document}